\documentclass[aps,prl,twocolumn,showpacs]{revtex4}
\usepackage{amsfonts}
\usepackage{amsmath}
\usepackage{amsthm}
\usepackage{graphics}
\usepackage{graphicx}
\usepackage{subfigure}
\usepackage{amssymb}
\usepackage{graphics}
\usepackage{graphicx}
\usepackage{color}
\usepackage{subfigure}

\def\kmax{k_{\rm max}}

\begin{document}

\title{Reply to ``Comment on ``Dispersive Bottleneck Delaying Thermalization of Turbulent Bose-Einstein Condensates'' by E. Kozik'' }
\author{Giorgio Krstulovic}
\affiliation{Laboratoire Cassiop\'ee, Observatoire de la C\^ote dÕAzur, CNRS, Universit\'e de Nice Sophia-Antipolis, Bd. de l'Observatoire, 06300 Nice, France}
\author{Marc Brachet}
\affiliation{Laboratoire de Physique Statistique de l'Ecole Normale 
Sup{\'e}rieure, \\
associ{\'e} au CNRS et aux Universit{\'e}s Paris VI et VII,
24 Rue Lhomond, 75231 Paris, France}
\date{\today}

\maketitle
%intro
{\bf Krstulovic and Brachet reply:} In the preceding Comment \cite{Kozik}, Kozik 
raised a criticism against the bottleneck 
proposed in our Letter \cite{Krstulovic2011PRLBottleneck} to be causing
a thermalization delay
when dispersive effects, controlled by the coherence length $\xi$,
are large at truncation wavenumber: $k_{\rm max}$.
The late-time energy spectrum presents
a front at wavenumber $k_c(t)$ propagating toward higher
wavenumbers and leaving in its wake a quasi-thermalized distribution.
Kozik argues that our observations agree with the relaxation scenario, 
developed by Svistunov  \cite{Svistunov}, 
that involves no bottleneck and predicts $k_c(t) \sim t^{1/4}$.

\begin{figure}[htbp]
\begin{center}
\includegraphics[width=0.4\textwidth]{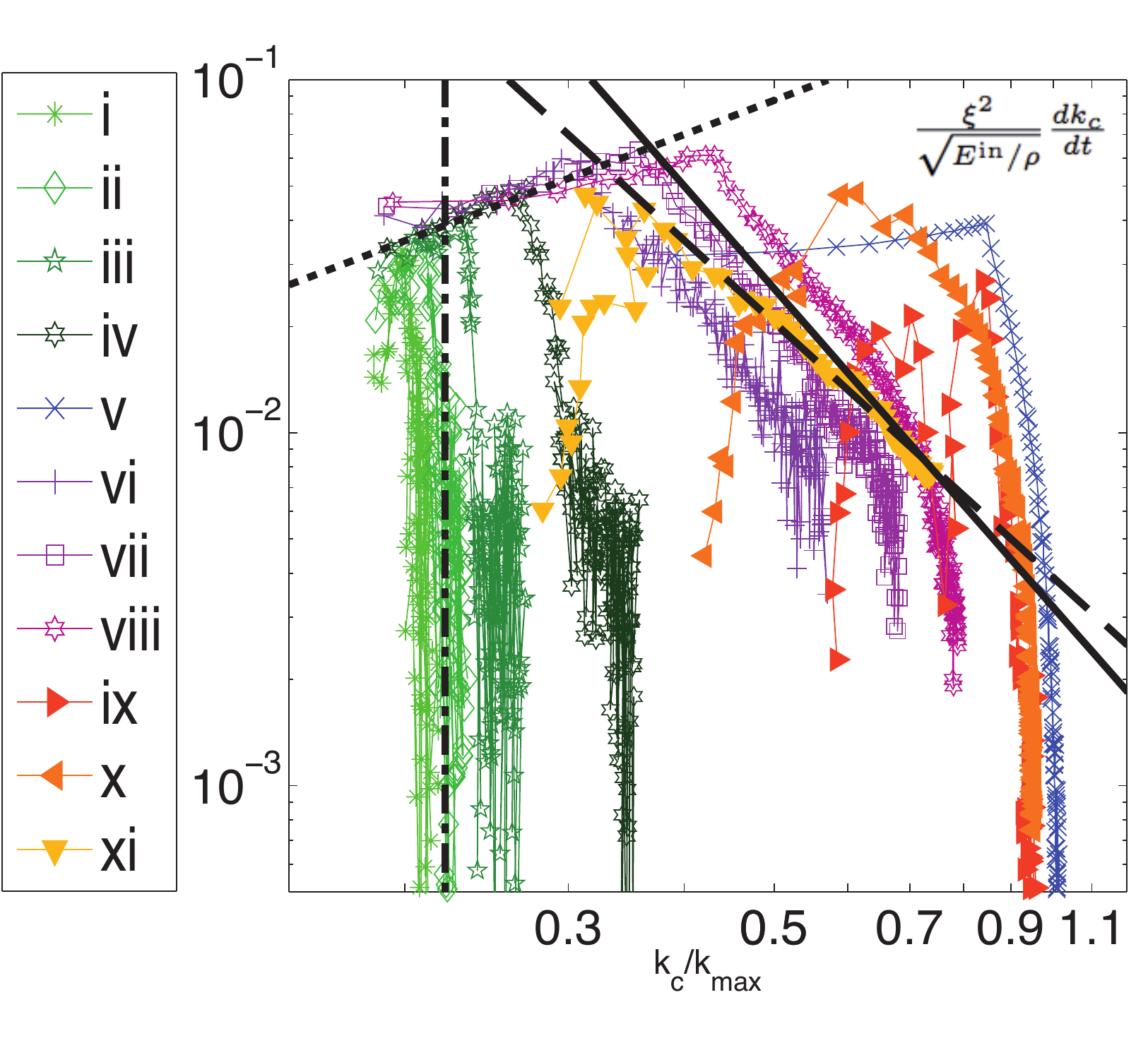}
\caption{(color online). Parametric representation $dk_{c}/dt$ v.s. $k_{c}/\kmax$ (adapted from Fig.4.e of \cite{Krstulovic2011PRLBottleneck}). Different scaling laws are displayed. Svistunov prediction: solid line $(k_c/k_{\rm max})^{-3}$; fit to run $xi$: large dashed line $(k_c/k_{\rm max})^{-2.4}$; dotted line $(k_c/k_{\rm max})^{1}$ and dotted-dashed line $(k_c/k_{\rm max})^{-\infty}$.}
\label{Fig}
\end{center}
\end{figure}

%Indeed
Indeed, it is apparent on Fig.\ref{Fig} (where
$k_c(t)\sim t^\alpha$ corresponds to a line of slope $(\alpha-1)/\alpha$) that four out of eleven runs ($vi$, $vii$, $viii$ and $xii$)
are somewhat compatible with the Svistunov prediction.
%
%However
However, the prediction only works in the limited range $0.4<k_c/k_{\rm max}<0.8$
where run $xi$, with Taylor-Green initial data and $\xi \kmax \sim 6$, 
yields a slope of  $-2.4$.
Runs $vi$, $vii$ and $viii$,  with initial data prepared using the stochastic Ginzburg-Landau equation and $\xi \kmax \sim 6$,
have slopes closer to the Svistunov prediction of $-3$.
Runs that saturate with $k_c/k_{max}\sim 1$ are reaching (truncated) thermal equilibrium and are not spectrally well-converged.
% In contrast
In contrast, runs $i$-$iv$, with $\xi \kmax \sim 24$,
saturate at $k_c/k_{\rm max}<0.4$ and
are well-converged  but the data suggests a
logarithmic growth of $k_c(t)$ (vertical line on Fig.\ref{Fig}), a behavior very different from that predicted in \cite{Svistunov}.

This discrepancy is perhaps due to the fact that Svistunov considers a two stages process: first a condensation produced by a particle-flux wave propagating to low energies and then a wave propagating from the low to high energy region. 
It is not absolutely clear that
the initial conditions of our Letter \cite{Krstulovic2011PRLBottleneck},
really correspond to any of the stages considered by Svistunov (see the discussion following Eq. (4.7) of \cite{Svistunov}).

Concerning the criticism against our use of the word "bottleneck", 
we believe it is related to a limitation in Svistunov theory. 
Indeed, it is well known that Bogoliubov's dispersion relation 
$\omega_{\rm B}(k)=k c (1+k^2\xi^2/2)^{1/2}$ (where $c$ is the sound velocity)  
implies (around wavenumber $k \sim 1/\xi$)
a change from propagative to dispersive behavior. 
This elementary point is not completely addressed in Svistunov theory,
in particular at level of the kinetic equations 3.10-3.13 of \cite{Svistunov} 
and Eq.(1) of \cite{Kozik}
\footnote{The correct collision integral that takes into account the Bogoliubov
dispersion relation is 4-26 of \cite{Dyachenko199296}.}.
Thus Svitunov's analysis is only applicable for wavenumbers $k>>1/\xi$. 
This limitation does not allow one to appreciate the importance of $\xi$ and to grasp that 
$k \xi$ (in particular $\xi k_{max}$) is an important dimensionless parameter
in this problem leading to a crossover between different regimes (see Fig.\ref{Fig} and also Fig.7b-c of \cite{Krstulovic2011PRE_TGPE}).

%Beliaev-Landau
In a physical BEC, $\kmax$ correspond to the equipartition wavenumber $k_{\rm eq}$ (see \cite{Krstulovic2011PRLBottleneck} and Sec. IV of \cite{Krstulovic2011PRE_TGPE}).
Sinatra and Castin  \cite{SinatraCastin} have shown that the slowdown of thermalization reported in \cite{Krstulovic2011PRLBottleneck} can be related to the behavior of the (classical) damping rate around equilibrium that reaches a maximum around $k \xi\sim 3$ and decays for  $k \xi>>1$. 
They have established that, at fixed $k \xi$ well beyond its maximum, the (quantum) Beliaev-Landau damping rate approaches the classical one provided $ k_B T/|\hat{\psi}_{\bf 0}|^2 g > 200$ which could be achieve experimentally using Feshbach resonance.
%conclusion

%In summary, we think that the theory of \cite{Svistunov} does not apply to our Letter as it fails to predict the behavior of runs $i$-$iv$, in the precise limit of large $\xi k_{max}$ and small $k_c/k_{\rm max}$ where the theory should be better applicable.

We acknowledege helpful discussion with S. Nazarenko.

Giorgio Krstulovic$^1$ and Marc Brachet$^2$.

{\small$^1$Laboratoire Cassiop{\'e}e, Observatoire de la C\^ote dÕAzur, CNRS, Universit\'e de Nice Sophia-Antipolis, Bd. de l'Observatoire, 06300 Nice, France.

$^2$Laboratoire de Physique Statistique de l'Ecole Normale 
Sup{\'e}rieure, 
associ{\'e} au CNRS et aux Universit{\'e}s Paris VI et VII,
24 Rue Lhomond, 75231 Paris, France.}

%\bibliographystyle{apsrev}
%\ibliography{BibReply}
%\begin{thebibliography}{99}

\end{document}